\documentclass[smallextended,final,natbib]{svjour3}
\usepackage{epstopdf}
\usepackage{graphicx}
\usepackage{rotating}
\usepackage{subfigure}
\usepackage{url}
\usepackage{siunitx}
\usepackage{nicefrac}
\usepackage{pbox}
\journalname{Experimental Astronomy}
\begin{document}
\title{Development of a Lunar Scintillometer as part of the National Large Optical Telescope Site Survey}
\titlerunning{Development of LuSci as part of the NLOT Site Survey}

\author{Avinash Surendran \and Padmakar S. Parihar \and Ravinder K Banyal \and Anusha Kalyaan}
\institute{Avinash Surendran \email{asurendran89@gmail.com} \and Padmakar S. Parihar \and Ravinder K Banyal
\at Indian Institute of Astrophysics, Bangalore - 560034, India \and 
Anusha Kalyaan \at
School of Earth and Space Exploration, Arizona State University, Tempe, AZ 85281, United States}

\date{Received: 13 April 2017 / Accepted: 18 December 2017}

\maketitle

\begin{abstract}
Ground layer turbulence is a very important site characterization parameter used to assess the quality of an astronomical site. The Lunar Scintillometer is a simple and effective site-testing device for measuring the ground layer turbulence. It consists of a linear array of photodiodes which are sensitive to the slight variations in the moon's brightness due to scintillation by the lower layers of the Earth's atmosphere. The covariance of intensity values between the non-redundant photodiode baselines can be used to measure the turbulence profile from the ground up to a height determined by the furthest pair of detectors. The six channel lunar scintillometer that has been developed at the Indian Institute of Astrophysics is based closely on an instrument built by the team led by Andrei Tokovinin of Cerro Tololo Inter-American Observatory (CTIO), Chile \citep{2010MNRAS.404.1186T}. We have fabricated the instrument based on the existing electronic design, and have worked on the noise analysis, an EMI (Electromagnetic Induction) resistant PCB design and the software pipeline for analyzing the data from the same. The results from the instrument's multi-year campaign at Mount Saraswati, Hanle is also presented.
\end{abstract}


\section{Introduction}

The first attempt at linking atmospheric turbulence measurements with the covariance of solar
scintillation between all baseline pairs by \citet{seykora1993} involved the analysis of the effect of solar scintillation on the solar limb motion. \citet{beckers1997} demonstrated that the solar scintillation indices from the baselines of a linear array of sensors could be used to measure the vertical optical turbulence profile. Based on whether observations are to be done during the day or night and the nature of the vertical turbulence profile required, the scintillations from different extended astronomical sources can be used \citep{tokovinin2007}. \citet{hickson2004} used a linear array of photodiodes to measure the scintillation of the moon and extracted the vertical turbulence profile from the same. Lunar scintillometers have been used around the world extensively for site survey campaigns to understand ground layer atmospheric seeing \citep{thomas2008,villanueva2010,lombardi2013}. It is usually used in conjunction with conventional turbulence monitors like the Multi-Aperture Scintillation Sensor (MASS) and the Differential Image Motion Monitor (DIMM) to provide a complete turbulence profile \citep{thomas2012,berdja2011}. \citet{newman2012} has also used the instrument for characterizing the ground layer adaptive optics (GLAO) instrument at the Multiple Mirror Telescope (MMT).

Most of the atmospheric fluctuations that contribute to seeing occur in a turbulent boundary layer that is at or near ground level. By measuring the strength and thickness of the ground layer, one can determine the minimum height at which a telescope should be placed to avoid much of this source of interference. In a GLAO system, real-time turbulence profiling is necessary to conjugate the deformable mirror to a height which contributes most to the wavefront distortions \citep{tokovinin2004}. Since regular site monitors can overestimate the ground layer turbulence \citep{tokovinin2010}, a robust and portable ground layer seeing monitor was required for surveying the site. 

The Lunar Scintillometer was chosen as an inexpensive and portable ground layer seeing measurement device, which could be transported through a difficult terrain and used in a harsh environment with minimal resources. The primary objective was to create an easily replicable instrument with a robust software pipeline for the automated analysis of several nights of data, as part of a multi-year campaign. \citet{2010MNRAS.404.1186T} has designed a robust six channel lunar scintillometer at CTIO, along with the profile restoration method to extract a vertical atmospheric seeing profile. The basic electronic design and profile restoration method are based on the instrument built at CTIO, Chile. We have worked on the noise analysis, instrument validation, a PCB design to mitigate parasitic capacitance and an automated software pipeline to analyze the data from multiple nights.

The construction of a large optical/NIR (Near Infrared) telescope in India was recommended by a committee on behalf of the Department of Science and Technology (DST) in 1986. The Himalayan Chandra Telescope (HCT), with a primary aperture diameter of 2 m, was to be the first step in gaining experience to build a bigger telescope at the extremely high altitude \citep{prabhu2014}. The telescope opened to the public in May 2003 and would form the stepping stone to construct a larger telescope in the same region. The National Large Optical Telescope (NLOT) is a proposed 8-10 m class telescope to expand the observational capabilities of Indian astronomers and to serve as a training ground for collaborative projects like the Thirty Meter Telescope (TMT). The Lunar Scintillometer is part of a multi-instrument campaign to ascertain the best site for building NLOT. Based on preliminary findings in the year 2007-08, Mount Saraswati at Hanle was chosen for a detailed site survey using multiple instruments \citep{parihar2015}.

Section~\ref{sec:TOP} explains the basic equations which govern the relationship between lunar scintillation and the vertical turbulence profile for the ground layer. In Section~\ref{sec:ID}, we describe the mechanical and electronic design of the instrument. Section~\ref{sec:TV} outlines the testing and validation techniques developed in the lab and performed on-sky, for ensuring the data quality of the instrument. The back-end software pipeline for data analysis is explained in Section~\ref{sec:SA} and the summary of the results from the complete two-year campaign are presented in Section~\ref{sec:RC}.

\section{Theory of Operation} \label{sec:TOP}

\begin{figure}
\centering
\includegraphics[width=1\textwidth]{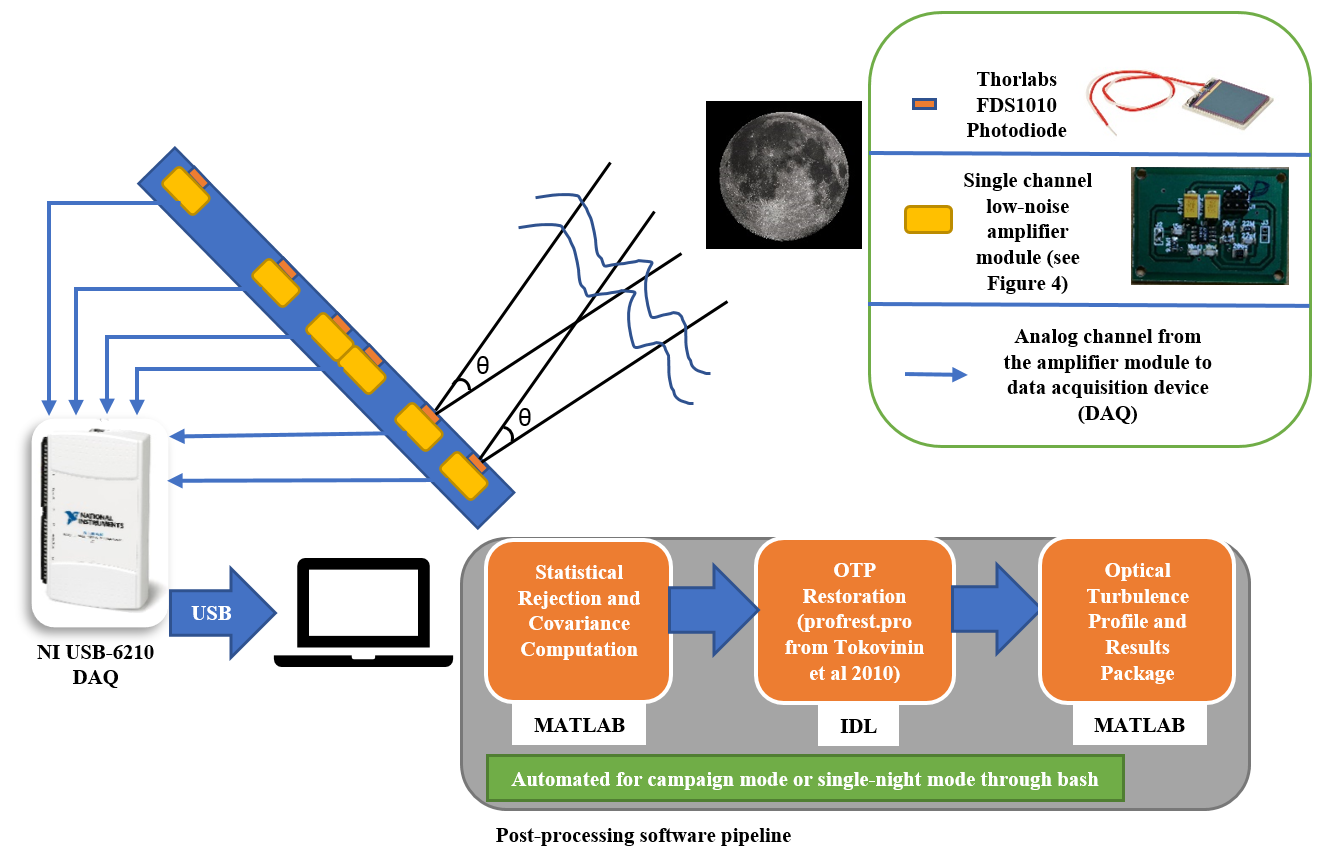}
\caption{Lunar Scintillometer - Principle of operation and software pipeline. \label{fig:LuSci}}
\end{figure}

The schematic of the instrument is shown in Figure~\ref{fig:LuSci}. We use a linear array of six photodiodes to measure the atmospheric induced instantaneous fluctuations of light coming from the moon. The photodiodes are arranged to form a set of non-redundant baselines along the same line. According to the height of the atmospheric layers to be sampled, the baselines can be selected according to the formula, $\displaystyle{z=\frac{r}{\theta}}$, with a resolution of $\displaystyle{{\Delta}z=\frac{d}{\theta}}$. Here, $r$ is the length of a baseline, $\theta$ is the vertex angle of the light cone (angular diameter of the moon), $d$ is the detector diameter and $z$ is the height at which the optical turbulence profile is to be measured. The covariance of the intensity fluctuations at the different baselines is used to measure the vertical profile of atmospheric turbulence in the lower layers of the atmosphere. We sample the photocurrents generated by each photodiode at an interval of every 2 ms through a data acquisition device. For every minute of observation corresponding to 30,000 samples of the photocurrent, the normalized covariances between the 15 different baselines are computed. The normalized covariance is given by the formula,
\begin{equation} \label{eq:scindex}
B(i,j)=\frac{1}{N}\sum_{n=1}^{N}(\zeta_i\zeta_j)_n,
\end{equation}
where $\displaystyle{\zeta_i=\frac{I_i}{\langle{I_i}\rangle}-1}$ is the normalized fluctuation of the photocurrent at the $i^{th}$ detector and N is the number of signal samples collected over the accumulation time of one minute.

The covariances are converted to the distribution of atmospheric refractive index constants, at configurable pivot heights along the line of sight. The relationship between the measured covariance (along a baseline, $r$) and the refractive index constant $C_n^2$ is given by \citep{2010MNRAS.404.1186T},
\begin{equation} \label{eq:wf}
B(r)=\int_{0}^{\infty}W(r,z)C_n^2 (z)dz,
\end{equation}
Here, $z$ is the propagation path from the moon to the instrument, $r$ is the baseline between the different photodiodes and $W(r,z)$ refers to the weighing function (WF) calculated from the instrument parameters, the phase of the moon, and the turbulence model. The WF is measured in units of m$^{-1/3}$. For the linear profile restoration, the weighing function is defined as a matrix of dimensions $B \times M$, where $B$ is the number of measured covariances from the baselines and $M$ represents a logarithmic distance grid of sampling points of $z$ (which are sufficient to capture the changes in the WF with $z$). The integral in equation~\ref{eq:wf} is thus substituted by a matrix multiplication. This method is used by \cite{thomas2008} for profile restoration. The profile restoration used by \cite{2010MNRAS.404.1186T} consists of a model which accounts for the moon's phase, the zenith angle and the baseline orientation. In the model fitting method, a model consisting of the $\log _{10} C_n^2$ values at specified pivot points (3 m, 12 m, 48 m, 192 m and 768 m in our case) is created with the values in between the pivot points estimated by linear interpolation. This is same as the separation of $C_n^2$ on a set of triangular functions with its vertices at the pivot points. The resultant values of $C_n^2$ are coarse, and a more refined result is obtained by non-linear fitting of a complex model, which also ensures that the value of $C_n^2$ is always positive. The model covariances computed from the set of $\log _{10} C_n^2$ values are compared with the measured values to minimize the error in the model with a least-square error minimization mechanism. The mathematical equations pertaining to the model fitting method and the linear restoration method can be found in \cite{2010MNRAS.404.1186T} and \cite{tokovinin2008}.
\section{Instrument Design}  \label{sec:ID}

\subsection{Mechanical Design} \label{subsec:MD}

\begin{sidewaysfigure}
\vspace*{+4.5in}
\includegraphics[width=1\textwidth]{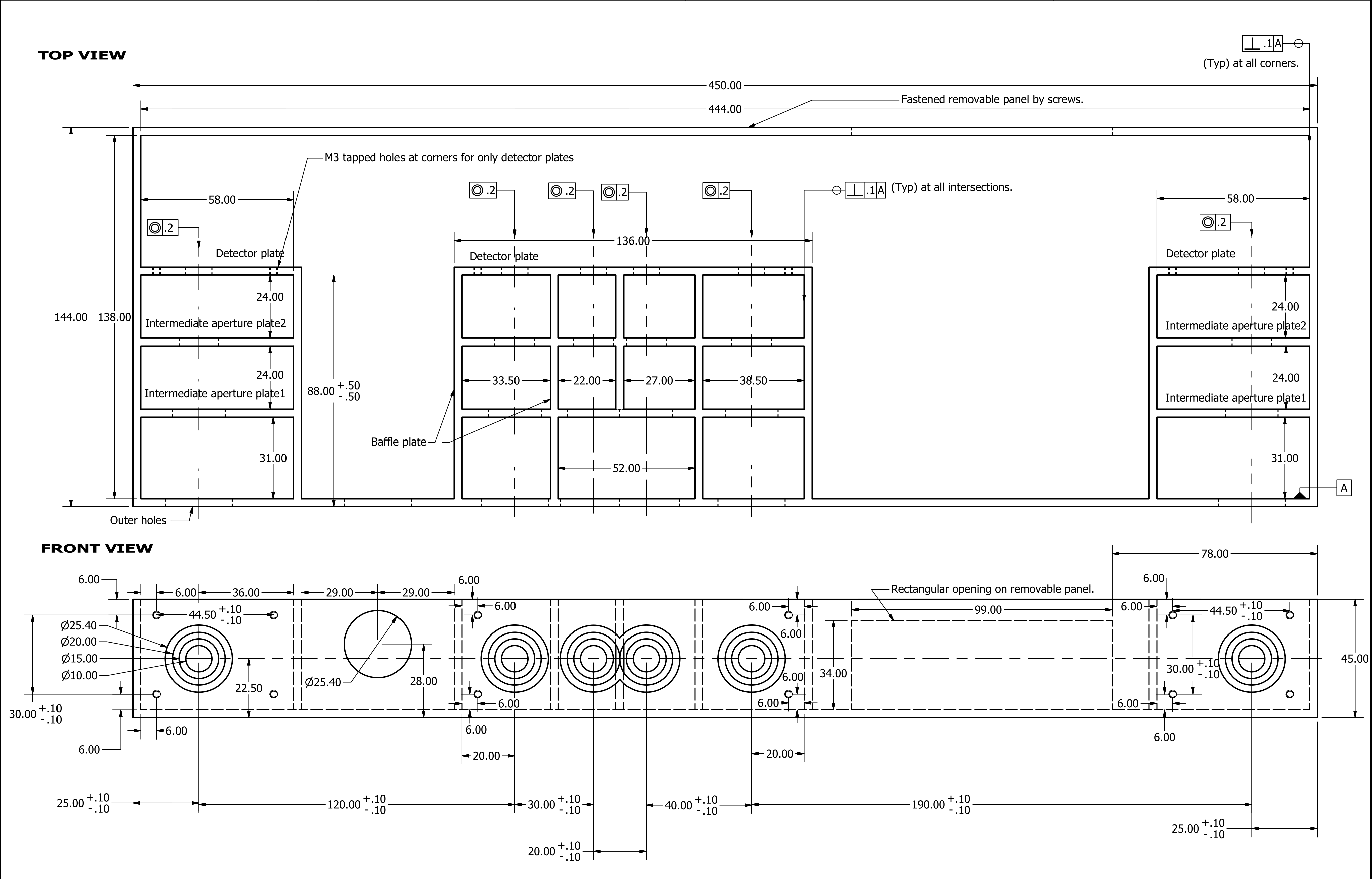}
\caption{Mechanical Design \label{fig:mdesign}}
\end{sidewaysfigure}

\begin{figure}
\centering
\includegraphics[width=1\textwidth]{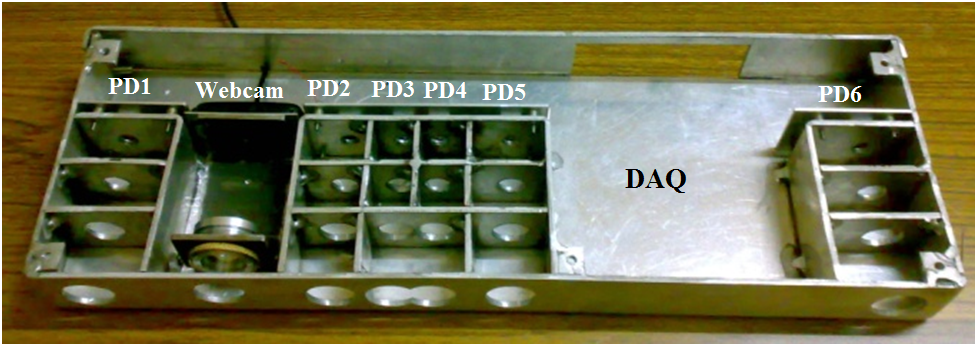}
\caption{Fabricated mechanical housing for detector modules of the Lunar Scintillometer and the web camera (for verifying that the moon is in the field-of-view of observation of the instrument). The positions of the webcam, the data acquisition device (DAQ) and the six photodiodes (PD1 - PD6) are shown in figure.
\label{fig:mfabrique}}
\end{figure}

Six photodiodes and their associated electronics are enclosed inside an aluminum box having the dimensions, 450 mm $\times$ 144 mm $\times$ 45 mm (as shown in Figure~\ref{fig:mdesign}). The position of the detectors from the first detector are (0, 12, 15, 17, 21, 40) cm, which contributes to 15 non-redundant baselines varying from 2 cm to 40 cm. Additional mounting plates are provided for the sensors and their associated electronics for each channel along with supporting clamps. A circular aperture of 1 cm diameter is the entrance aperture for each photodiode. Although no optics is used to concentrate the light on the photodiode, baffling provides an un-vignetted field of 10$^{\circ}$ for each photodiode. The individual photodiodes are isolated from each other by walls, except for the closest pair (baseline of 2 cm) where the outer holes partially overlap. The enclosure is anodized and painted black to prevent stray light reflections. The fabricated aluminum box (prior to anodizing) is shown in Figure~\ref{fig:mfabrique}.

\subsection{Electronic Design} \label{subsec:ED}

\begin{figure}
\centering
\includegraphics[width=0.8\textwidth]{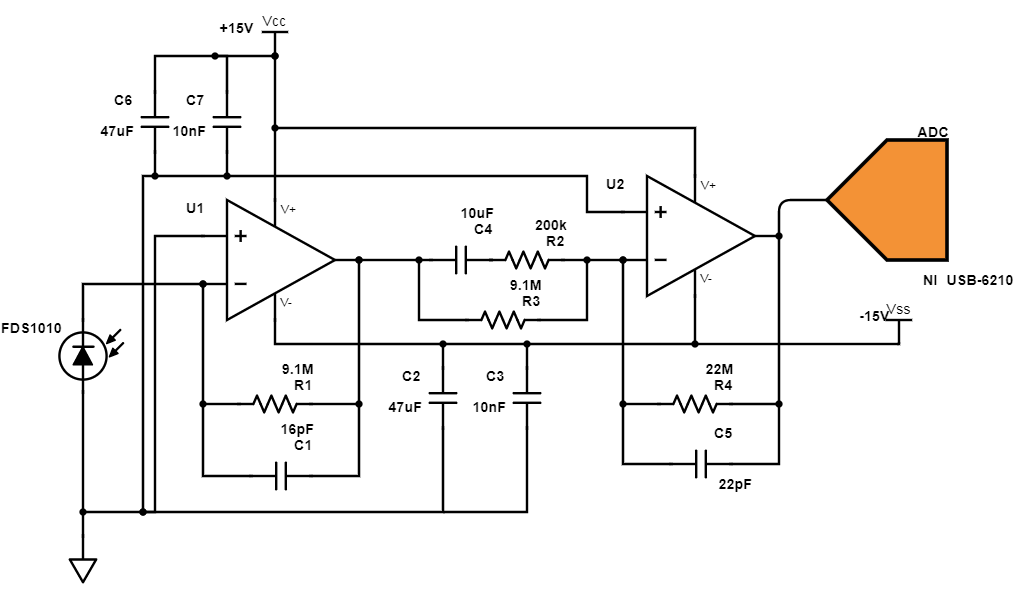}
\caption{Electronic design of the amplifier module. U1 and its associated elements form a high-gain transimpedance amplifier, and U2 and its associated elements form an amplifier which amplifies DC and AC voltages with different gains. \label{fig:edesign}}
\end{figure}

The FDS1010 \citep{fds1010}, a silicon photodiode, is used as the primary photodetector of the scintillometer. It has a square active area of $10 \times 10$ mm$^2$. At a bias voltage of \mbox{-5 V}, the dark current generated by the photodiode would amount to about \mbox{600 nA}. So, the photodiode is operated in the unbiased (photovoltaic) condition, where it generates negligible dark current. It generates a direct photocurrent of 52.36 nA (see Appendix~\ref{app:fmoon}) for the full moon, and a constantly varying current of amplitude as small as $10^{-4}$ times the direct current. The purpose of the electronics is to convert this current to a measurable value, adding minimal electronic noise in the process.

\begin{figure}
\centering
\includegraphics[width=1\textwidth]{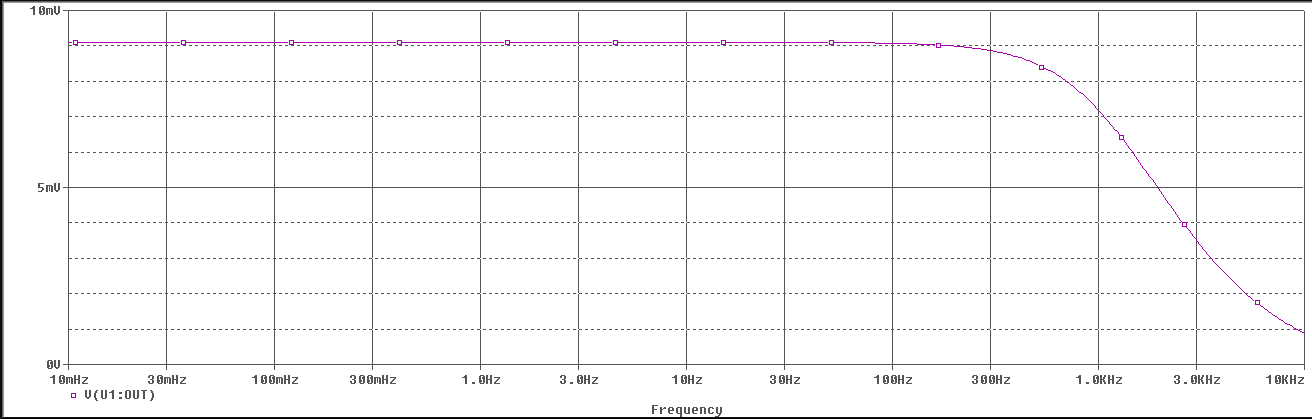}
\caption{Simulated frequency response of transimpedance amplifier (x-axis shows frequency in hertz and y-axis shows amplitude in millivolts). \label{fig:freq1}}
\end{figure}

The complete electronic design is given in Figure~\ref{fig:edesign}. The current is amplified and converted into voltage by a low noise transimpedance amplifier (U1) with a feedback resistance of 9.1 M$\Omega$ and a compensation capacitance of 15 pF to prevent oscillations and to limit the 3dB upper cut-off frequency to 1.29 kHz. The frequency response of the transimpedance amplifier design is simulated with the photodiode, which is assumed to be a constant current source in parallel with an internal capacitance of 1243 pF \citep{fds1010}. The frequency response of the transimpedance amplifier is shown in Figure~\ref{fig:freq1}.

As the fluctuations in the photocurrent caused by scintillation can be as low as $10^{-4}$ times the direct current generated by the full moon, direct digitization of the signal with 16-bit resolution would be barely sufficient. An inverting amplifier (U2) separates the variable (AC) part of the signal (due to scintillation) from the average (DC) value (due to the brightness of the full moon). The variable part of the signal, which is extracted by a high-pass filter with a cut-off frequency of 0.1 Hz, is amplified by 90 times. The DC voltage, which is amplified by approximately two times, is also important as it plays a significant role in computing the normalized covariance. The ratio of the AC to DC voltage amplification factor is given by $K_{ampl} = 45$. Overall, the electronic design selectively amplifies the signal with a frequency greater than 0.1 Hz by $K_{ampl}$ compared to the signal less than 0.1 Hz. The amplified signals from the six channels are sent to a data acquisition device (DAQ) device, which is housed in the same box. The DAQ device reads the six channels sequentially at a rate of 5 kHz and is median averaged every ten samples to acquire the signal at 500 samples/second.

A regulated $\pm$15 V dual output source with a maximum power of 2 W is used to power the detector modules. The power converter is sufficient enough to drive the electronics of the detector module, and the system is designed to operate with maximum portability. Any 12 V, 1 A DC power source can be used for powering the device. PTFE (Poly-Tetra-Fluoro-Ethylene) cabling and non-electrolytic capacitors were used inside the enclosure to tolerate the sub-zero temperature conditions in the Himalayan landscape.

\section{Testing and Validation} \label{sec:TV}

\subsection{Noise Analysis} \label{subsec:NA}

{\par\sloppy
A comprehensive noise modelling of the circuit predicts a non-photon noise voltage of $36.71 {\mu}$V$/{\sqrt{\mathrm{Hz}}}$. The contribution of Johnson noise amounts to $35.9 {\mu}$V$/{\sqrt{\mathrm{Hz}}}$ because of the set of large resistances used in the amplifier circuit. Over a 250 Hz frequency range (limited by the sampling frequency of 500 Hz), the predicted output noise voltage is 580 ${\mu}$V. The predicted photon noise generated by the full moon current is about 1.68 mV at the amplifier output (see Appendix~\ref{app:fmoon}). The non-photon noise is mostly dominated by Johnson noise.
\par}

\subsection{Lab Testing} \label{subsec:LT}

\begin{figure}
\hfill
\subfigure{\includegraphics[width=5cm]{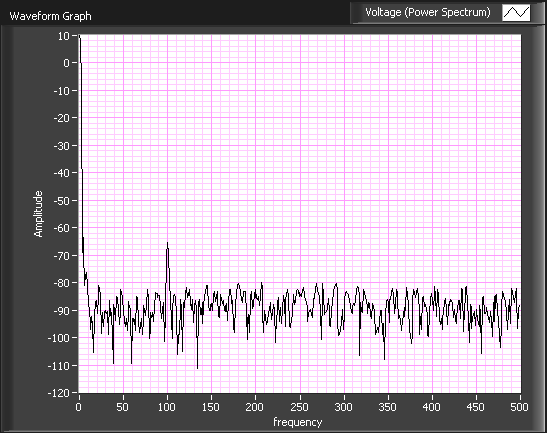}}
\hfill
\subfigure{\includegraphics[width=5cm]{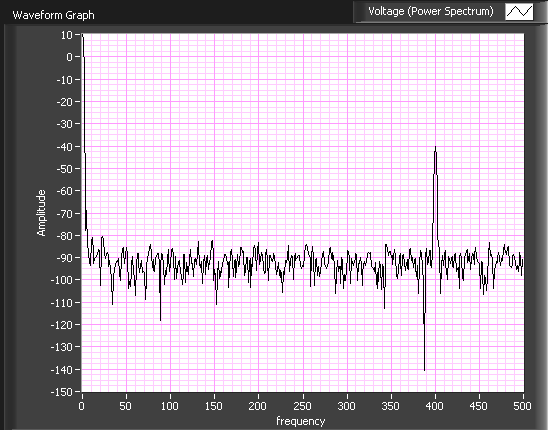}}
\hfill
\caption{Frequency response of the transimpedance amplifier when a test LED was subjected to (left panel) DC voltage of 4 V and an AC voltage of 0.0005 V at 100Hz (right panel) DC voltage of 4 V and an AC voltage of 0.01 V at 400 Hz. \label{fig:freq2}}
\end{figure}

Testing of the photodiode was done with light from a red LED (${\simeq} 630$ nm) in a dark room. The LED was subjected to a square-wave input with a DC amplitude of 4 V and a slew of AC amplitudes to evaluate the sensitivity of the photodiodes. The LED was also subjected to a variation of frequencies from 100 Hz to 400 Hz to evaluate the time response of the photodiode and the associated electronics. Results from the testing of the transimpedance amplifier output are shown in the form of screenshots of the amplitude spectrum taken in real-time (Figure~\ref{fig:freq2}). Amplitude is shown in decibels and frequency is in Hertz, in Figure~\ref{fig:freq2}.

In the first figure (Figure \ref{fig:freq2}, left), the difference between the DC and AC (100 Hz) voltage amplitude from the photodiode was measured to be 51 dB, which corresponds to a DC voltage amplitude ${\approx}350$ times the AC voltage amplitude. In the second figure (Figure \ref{fig:freq2}, right), the difference between the DC and AC (400 Hz) voltage amplitude was measured to be 76 dB, which corresponds to a DC voltage amplitude ${\approx}$6,500 times the AC voltage amplitude. This conforms roughly with the ratio of the DC to AC voltage given to the red LED used for testing the photodiode.

When the circuit was assembled on a generic PCB (Printed Circuit Board) and tested in a dark room, a high amplitude sinusoidal output signal was obtained. The signal measures about 10-12 V with no DC offset and has a frequency of 564 Hz. When tested with a smaller feedback resistance of 700 k${\Omega}$ at the inverting amplifier, the oscillations disappear but it adds considerable 50 Hz noise to the output. In the current scenario, the saturated signal would swamp the photocurrent and would not allow the scintillations to be measured accurately.

\subsection{Parasitic Capacitance and Final Fabrication} \label{subsec:PC}

\begin{figure}
\hfill
\subfigure{\includegraphics[width=5cm]{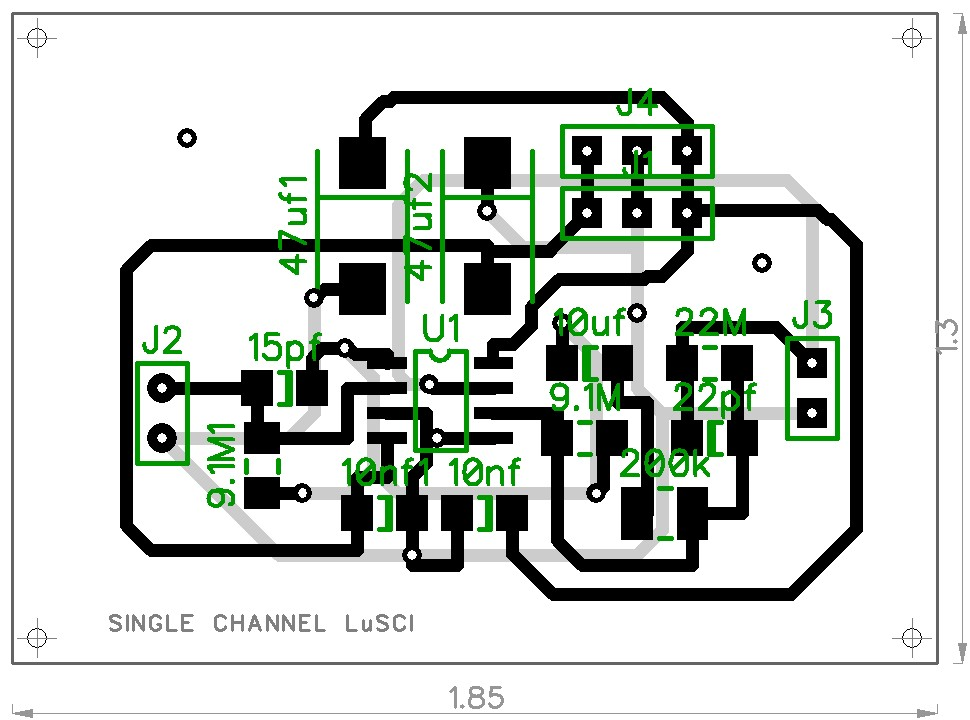}}
\hfill
\subfigure{\includegraphics[width=5cm]{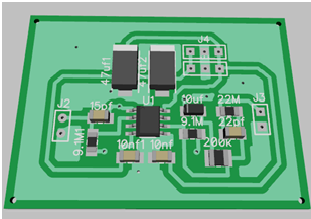}}
\hfill
\caption{(Left panel) Final PCB design for Lunar Scintillometer (without the ground plane) and the (right panel) 3D model of the PCB showing the component footprints. \label{fig:pcb_design}}
\end{figure}

The parasitic capacitance in a poorly designed PCB is of the order of a few pF and should not affect the output performance of most of the circuit configurations. In a stable amplifier having a high amplification factor within a single stage, oscillations can either be caused due to grounding problems or due to the presence of parasitic capacitance acting in a positive feedback loop \citep{karki20001}. The proximity of metallic components to certain parts of the circuit also reduced the amplitude of oscillations, leading to the hint that parasitic capacitances were causing the oscillations. The poles of the transimpedance amplifier and one of the poles of the second stage of the amplifier were close, rendering the circuit marginally stable. Simulations were used to confirm the presence, magnitude and effect of parasitic capacitance on the output signal. The oscillations were replicated with simulations with variable parasitic capacitance ($<$ 1 pF) between the output node and the photodiode. A new PCB was designed (as shown in Figure~\ref{fig:pcb_design}) with stringent requirements on the parasitic capacitances, the quality of components used and the ability to eliminate any 50 Hz electrical noise. The new PCB design eliminates the oscillations and produces a white noise floor conforming to the electronic noise calculations.

\subsection{Instrument characterization}

The goal of the instrument characterization before deployment was:
\begin{enumerate}
  \item To ensure that the amplitude of AC voltage fluctuations is within the predicted theoretical noise output in the absence of light,
  \item To ensure the absence of contamination by ambient electromagnetic sources, and
  \item To ensure that the DC voltage output from the photodiodes is close to the predicted theoretical DC voltage output when exposed to the full moon.
\end{enumerate}

\begin{figure}
\centering
\includegraphics[width=1\textwidth]{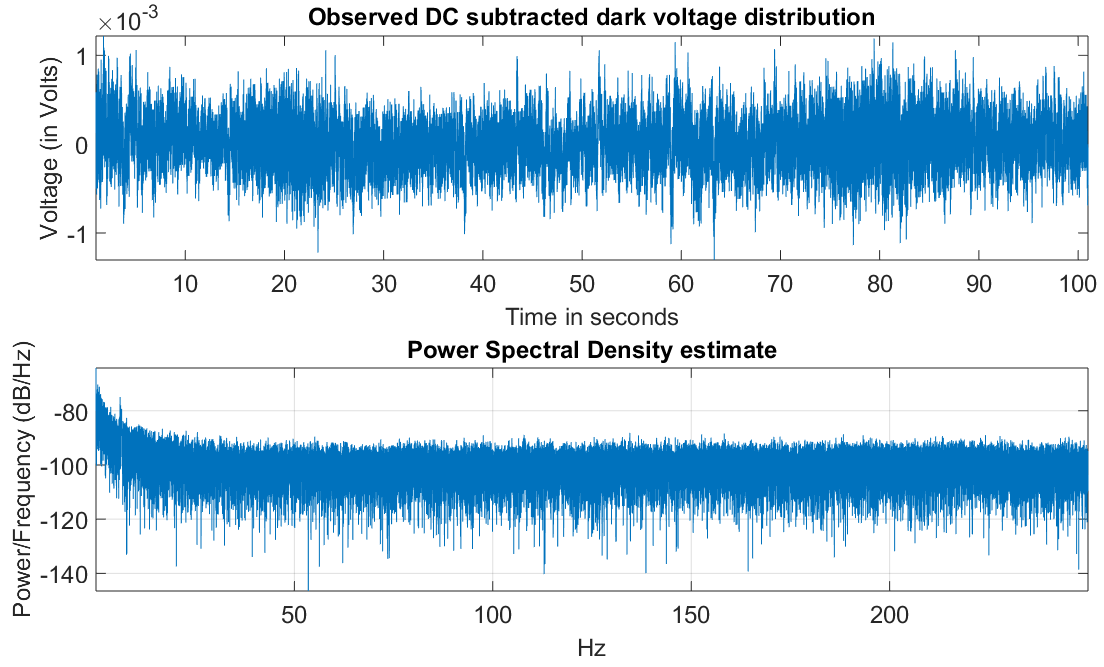}
\caption{Dark current output for photodiode number 5 (top panel), and its power spectral density (bottom panel). \label{fig:dark}}
\end{figure}

All the six photodiodes were tested in a dark room and the average noise output from the amplifier ranges from 400 to 800 ${\mu}$V for different photodiodes. Figure~\ref{fig:dark} shows a 200-second sample of dark current output of one of the photodiodes observed in the lab. The DC value of dark current is close to zero, and the RMS fluctuations amount to 643 ${\mu}$V for the current sample. The noise is close to the predicted noise output as outlined in Section~\ref{subsec:NA}. The output is also free of any ambient electromagnetic contamination as shown in the power spectral density plot in Figure~\ref{fig:dark}. The noise floor for  ${C_n^2}$ was found to be on average, 1-2 orders of magnitude less than that of observed ${C_n^2}$. The noise floor was estimated by adding a DC voltage of 1 V (DC voltage close to full moon) to the dark observations and running the data through the profile restoration algorithm.

From Section~\ref{app:fmoon}, the direct current generated by the full moon is given by 
\begin{equation}
I_{dc}=52.36 \times 10^{-9}A
\end{equation}
The ideal DC voltage output of each channel is given by
\begin{eqnarray}
V_{dc}&=&I_{dc} \times K_{trans} \times K_{dc} \nonumber \\
&=&52.36 \times 10^{-9}A \times 9.1 \times 10^{6} \Omega \times 2.417 \nonumber \\
&=&1.15 V
\end{eqnarray}
where $K_{trans}$ is the gain of the transimpedance amplifier (in ohms) and $K_{dc}$ is the DC gain of the inverting amplifier.

\begin{figure}
\centering
\includegraphics[width=1\textwidth]{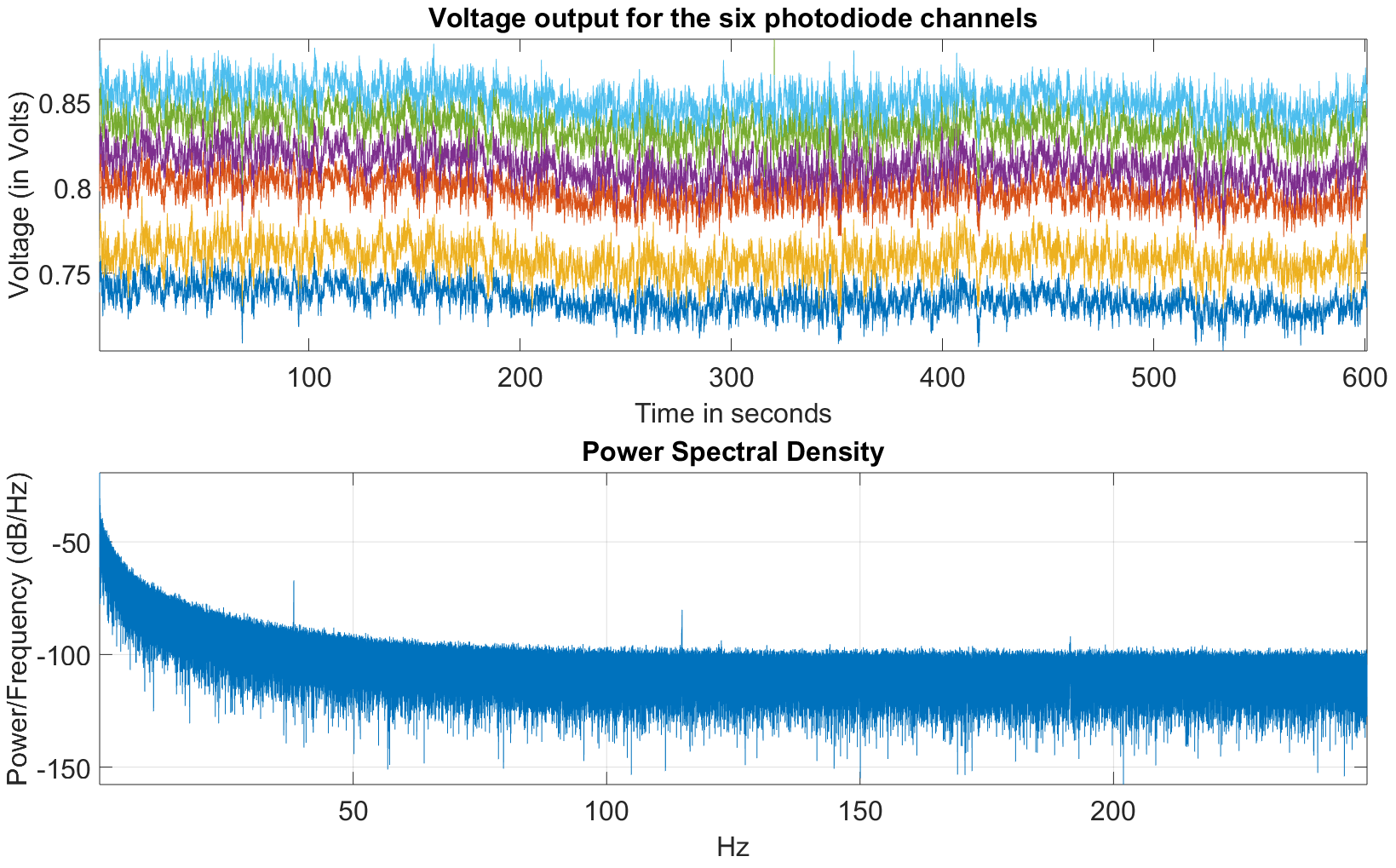}
\caption{Voltage output of all six photodiode channels when exposed to the full moon (top panel) and the power spectral density of the first channel (bottom panel). \label{fig:moonvolt}}
\end{figure}

\begin{figure}
\hfill
\subfigure{\includegraphics[width=5cm]{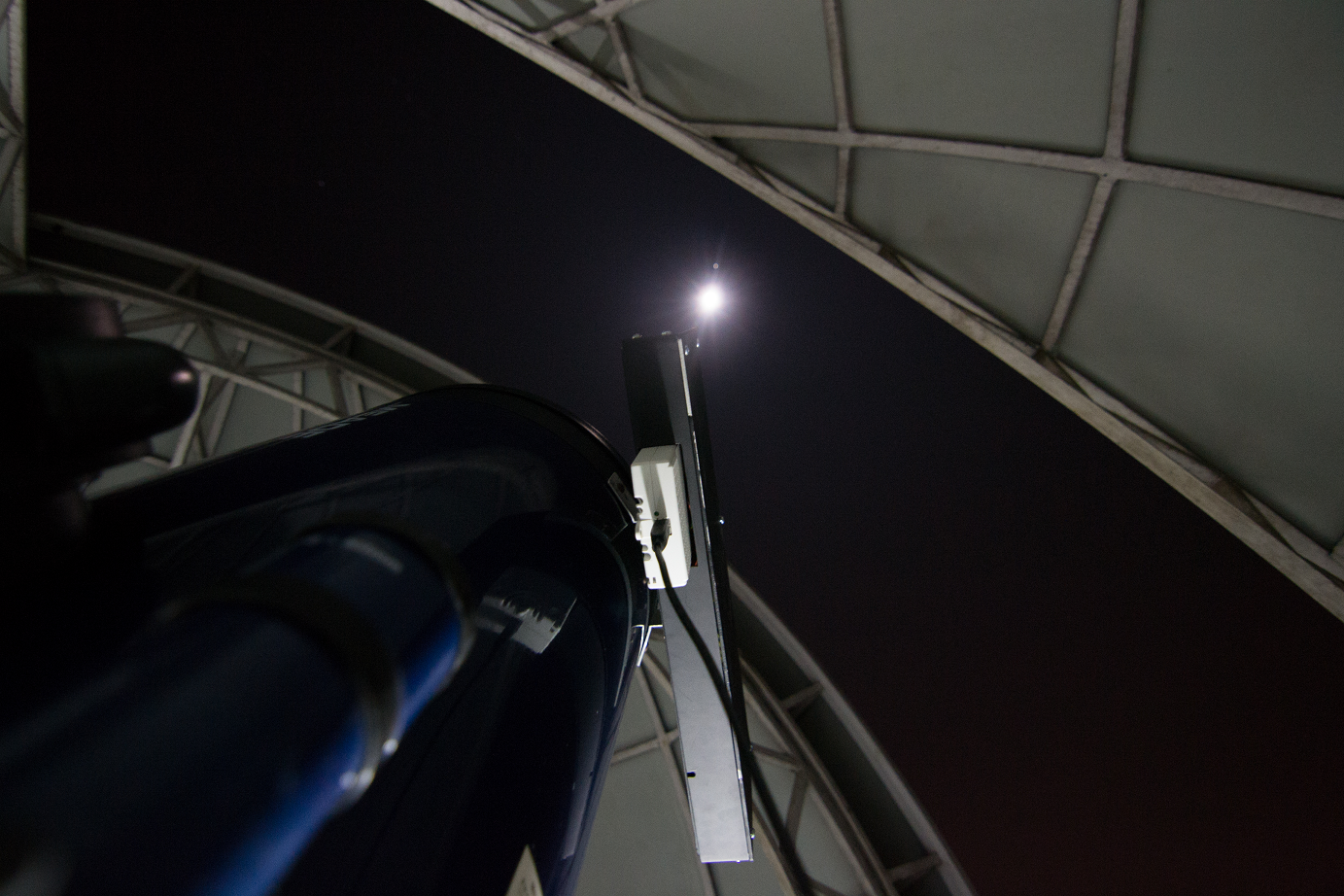}}
\hfill
\subfigure{\includegraphics[width=5cm]{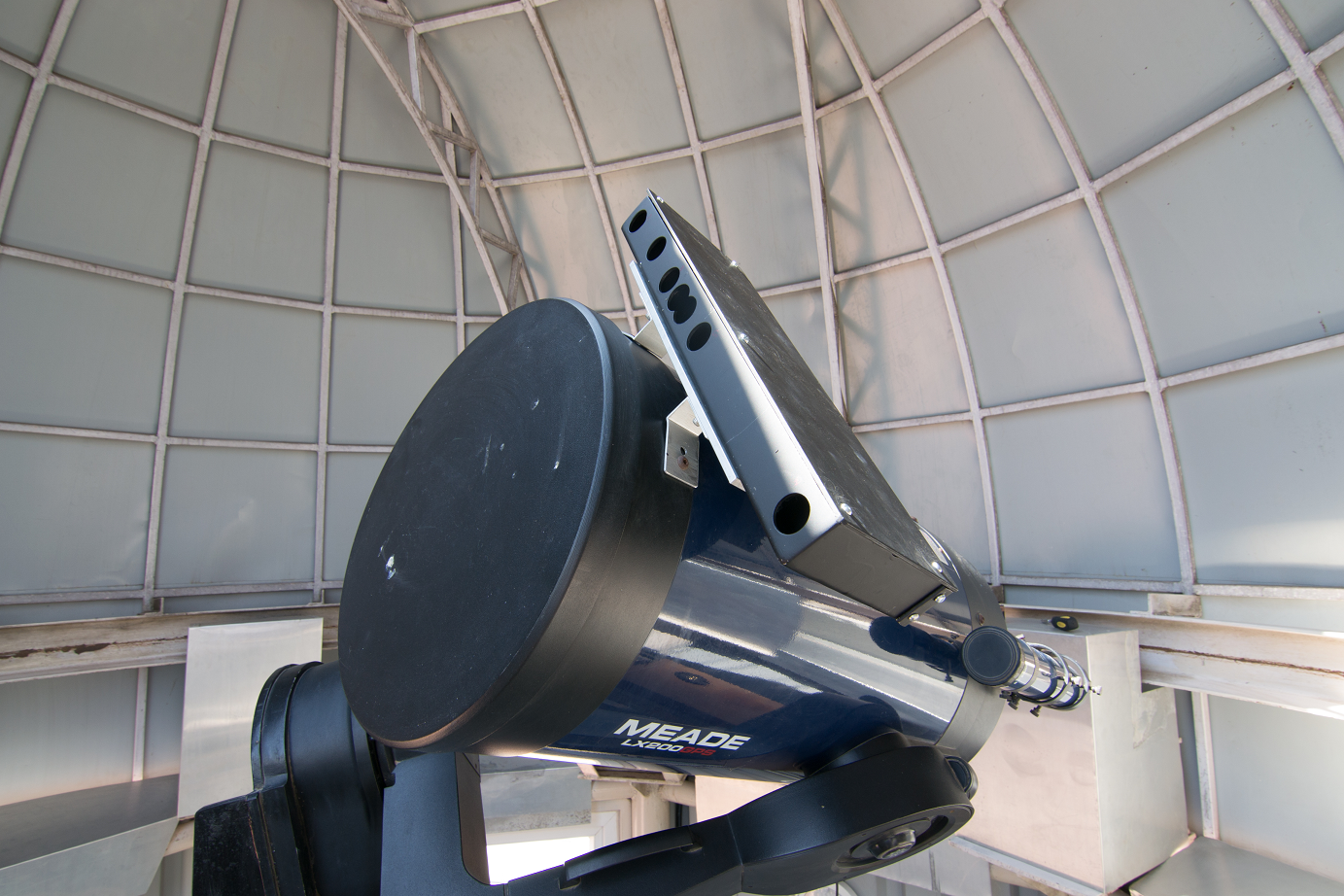}}
\hfill
\caption{(Left panel) The Lunar Scintillometer piggybacked on a Meade LX200 telescope tracking the moon (right panel). The front view of the Lunar scintillometer, showing the baffles through which moonlight reaches photodiodes. \label{fig:dome}}
\end{figure}

Figure~\ref{fig:moonvolt} (top panel) shows the output of the six photodiode channels when exposed to the full moon on 27th January, 2013 on a test run. Compared to the theoretically computed DC value of 1.15 V, the output voltage ranges from 0.73 V to 0.85 V for the six channels. The RMS fluctuations about the DC value for the output channels range from 7.7 mV to 9.2 mV. The reduction in the DC output voltage is due to various factors including the observation being done in a polluted urban environment and the deviation of the photodiode and circuit parameters from what is given in their respective datasheets. The variation of the output voltage between the photodiode channels is due to the difference in performance of the individual photodiodes and the components used in the amplifier circuit of the different channels. The computation of the normalized covariance compensates for such differences as can be seen from equation~\ref{eq:scindex}. Figure~\ref{fig:dome} shows the instrument piggybacked on a Meade LX200 telescope. All the local on-sky testing and characterization was done on this platform.

\section{Software and Analysis} \label{sec:SA}

The software to control and acquire data from the instrument is written in LabView and works on Windows. The software acquires data at 5 kHz and averages every ten samples to obtain the data at a target frequency of 500 Hz. The data is stored as a LabView Measurement file (LVM) with its respective headers. We rely on the pointing and tracking capability of the mount for tracking the moon. A webcam is provided to visually check if the moon is in the field of view and correct the field manually if need be. A combination of MATLAB, bash script and IDL on Ubuntu mounted in a portable virtual machine (VM) is used for the software pipeline.

The software pipeline is uploaded as a code repository on GitHub at \url{https://github.com/avinash-iiap/LuSci}. An accompanying manual for the software is also provided as a readme.md file in the repository. All files mentioned in this section under parentheses are available in the repository.

The software pipeline is divided into two parts:
\begin{enumerate}
  \item A stand-alone mode (executed by the script, lusci\_run1.sh) which processes a single LVM file from a single night, and creates the vertical turbulence profile for the same.
  \item A campaign mode (executed by the script, lusci\_run\_complete.sh) which processes all the LVM files in a specified folder, creates the vertical seeing distribution for each night and generates an overall analysis for the whole campaign.
\end{enumerate}

\subsection{Stand-alone mode} \label{subsec:SAM}

\begin{figure}[ht]
  \centering
  \includegraphics[width=0.9\textwidth]{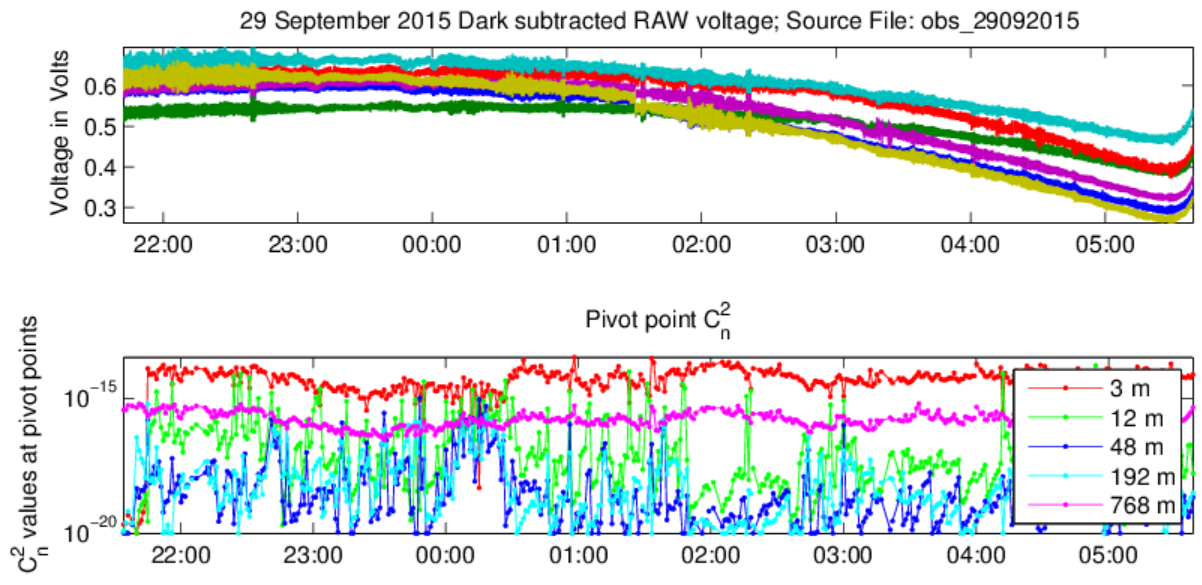}
  \caption{Voltage output (top panel) and the $C_n^2$ distribution (in m$^{\nicefrac{-2}{3}}$) for the night from all six photodiode channels when exposed to the full moon (bottom panel). \label{fig:voltage_cn2}}

  \vspace*{\floatsep}

  \includegraphics[width=0.8\textwidth]{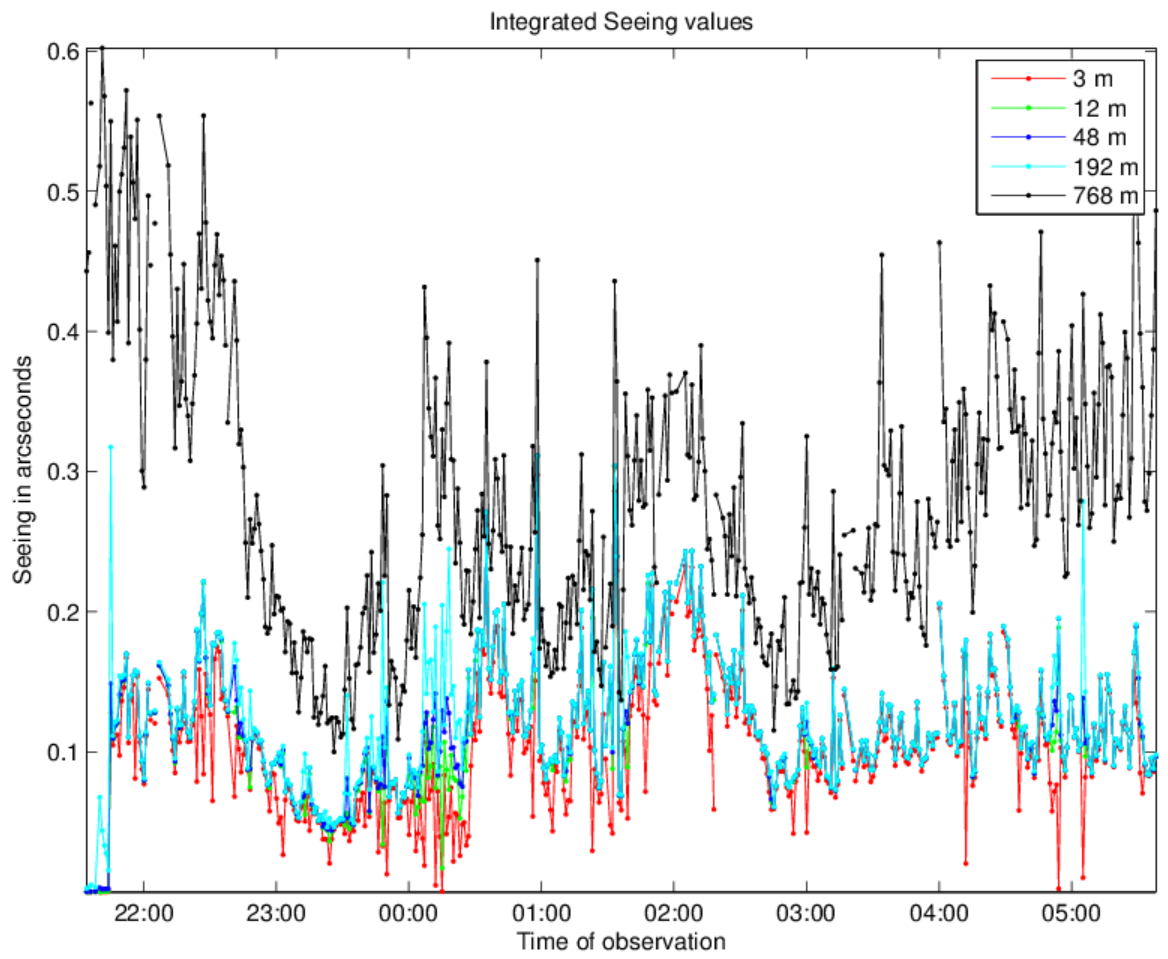}
  \caption{Integrated seeing (at a wavelength of 500 nm) on the night of 29th September 2015. \label{fig:Seeing}}
\end{figure}

\begin{figure}
\centering
\includegraphics[width=1\textwidth]{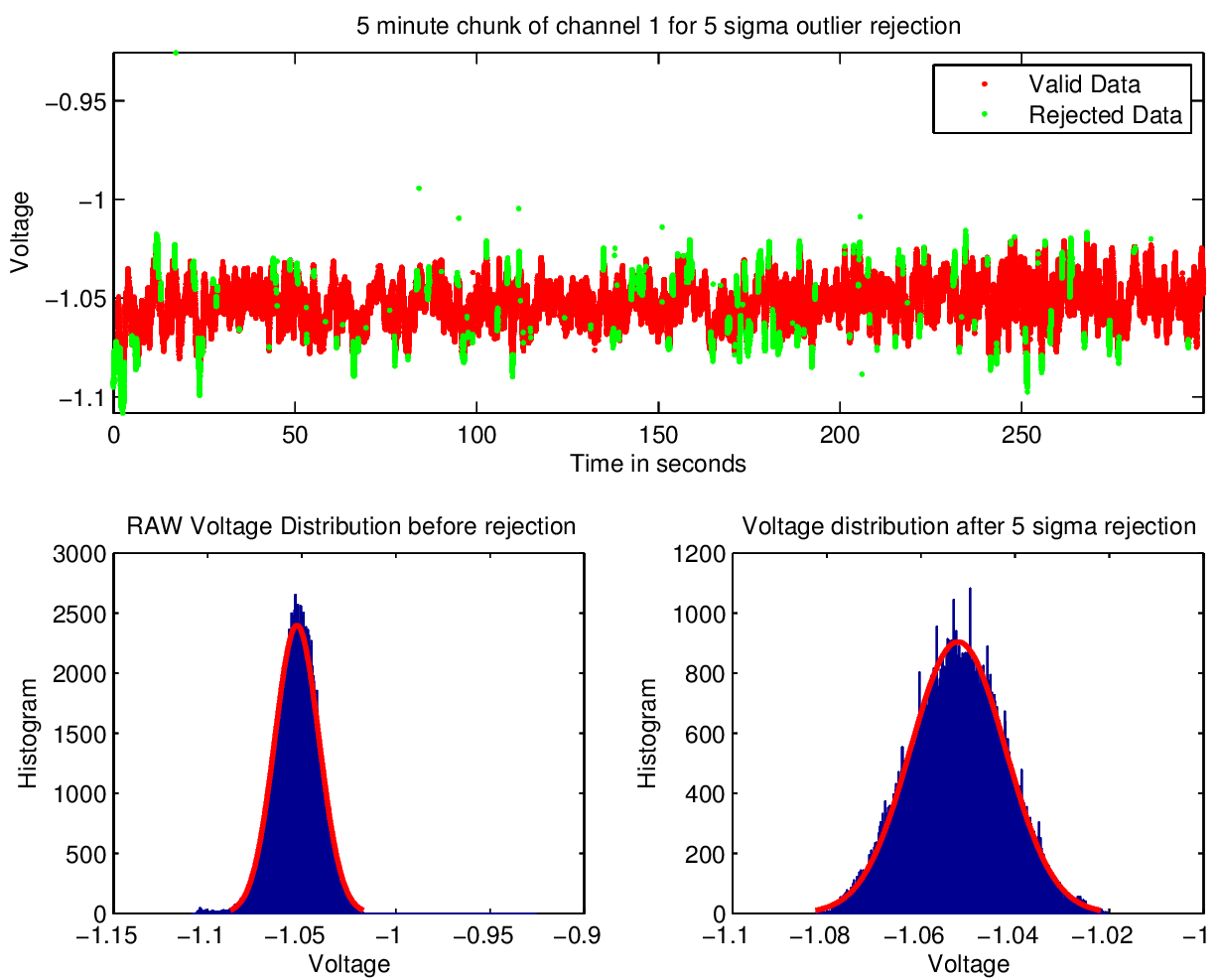}
\caption{A five minute chunk of voltage distribution for channel 1 (top panel), and the histogram before and after statistical elimination (bottom panel). \label{fig:voltage_sigma}}
\end{figure}

The input parameters to the stand-alone script are the number of voltage samples per second, latitude, longitude, detector positions on the linear array, accumulation time for computing the normalized covariances and the AC to DC amplification coefficient of the amplifier circuit. The stand-alone mode consists of the following scripts executed in the order given below:
\begin{enumerate}
{\par\sloppy
  \item A signal pre-processing package and a normalized covariance generator (voltage\_to\_covariance\_500samp\_v1.m).
  \item An IDL turbulence estimator program that converts the normalized covariances to the turbulence profile (profrest.pro).
  \item A plotting package which displays the turbulence statistics for each night (report\_compiled\_v1.m).
\par}
\end{enumerate}
The main challenge of the pre-processing package is to isolate the errors due to tracking, cloud opacity and vibrations in the telescope. Any data points above 2 V (which signifies a high light contamination) or below 0 V (which signify the absence of light or a data acquisition error) are rejected from the raw data. Owing to the random nature of atmospheric turbulence and its adherence to a normal distribution, the voltage data is then put through an iterative statistical elimination where any outliers outside of a five sigma normal distribution are eliminated. The elimination is done for every one-minute sample of 30,000 points. A normal distribution is fit on to every one-minute voltage distribution and the outliers outside of a five sigma limit are rejected. The iterative elimination gets rid of the data point corresponding to all six channels even if the data point of one of the channels is outside of the five sigma limit. This ensures that anomalies or spikes in the voltage due to pointing and tracking errors (which persist for a much shorter time than actual seeing variations) get eliminated, with a far less probability of actual underestimation of the turbulence integral. 

The normalized covariances are calculated according to equation~\ref{eq:scindex} using the instantaneous (AC) and averaged (DC) voltages over the accumulation time of one minute. The normalized covariances are written to a formatted text file which can be understood by the turbulence estimator. The estimation of the optical turbulence profile from the normalized covariances was done with an IDL turbulence estimator modelled on the profile restoration method given in \citet{tokovinin2008}.

The plotting package creates an array of plots which will be useful to the user in understanding the nature of turbulence and the performance of the instrument throughout the night. As an example, the data from 29th September 2015 is plotted. The optical turbulence plots consist of the filtered voltage distribution, ${C_n^2}$ distribution at the pivot point altitudes (Figure~\ref{fig:voltage_cn2}) and the integrated atmospheric seeing values from the ground level till the pivot point altitudes (Figure~\ref{fig:Seeing}). The results of the iterative elimination on the voltage distribution is shown in Figure~\ref{fig:voltage_sigma}. The power spectrum of all six channels are also computed.

\subsection{Campaign Mode}

The campaign mode wraps around the stand-alone mode. The campaign mode accepts the input folder where all the LVM files corresponding to the raw voltage data from different nights are located and executes the stand-alone analysis of each night. It executes the stand-alone analysis after setting the input parameters (required for the stand-alone analysis) automatically for each night of observation. The campaign mode script helps in the automation of the overall analysis of all the data that the user has and gives a comprehensive plot of the variation of the optical turbulence profile during the period in which the user has taken the data.

\section{Results and Discussion} \label{sec:RC}

\begin{figure}
\centering
\includegraphics[width=1\textwidth]{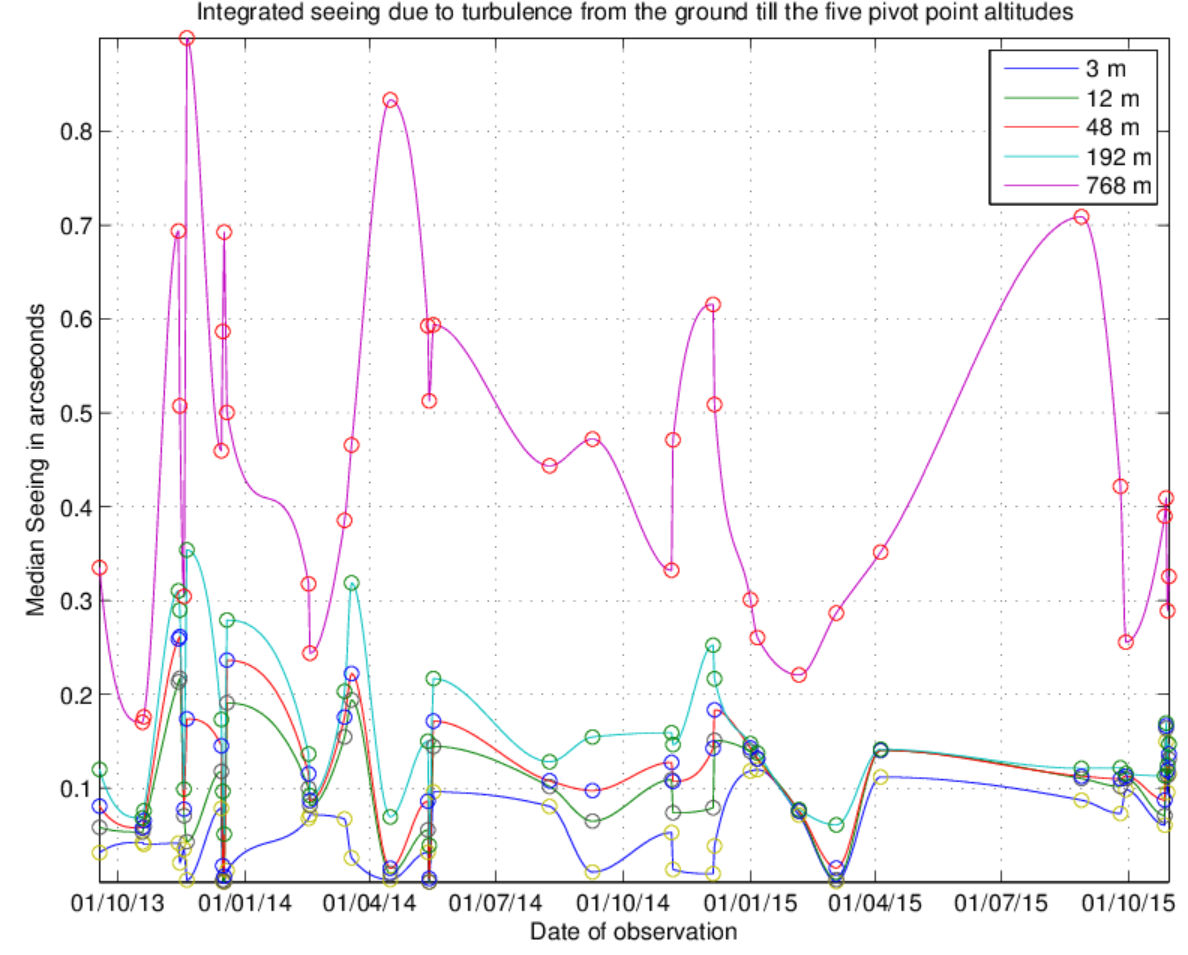}
\caption{Atmospheric seeing distribution (at a wavelength of 500 nm) from September, 2013 to October, 2015. \label{fig:compiled_seeing}}
\end{figure}

The target of the ground layer seeing campaign was to ascertain the strength of turbulence in the ground layer and the pattern of variation of the same at the site. The observations were taken for a 5 - 7 day period centered around the full moon each month, depending on the quality of the sky. The observations were conducted as part of a two-year campaign from September 2013 to October 2015 at the same site. The technique of thresholding and iterative statistical elimination (as explained in Section~\ref{subsec:SAM}) was used to separate the nights where the data could be converted into useful data, from the nights where we could not (due to a combination of factors including persistent cloud cover, vibrations and pointing and tracking errors). Each set of voltage samples over a period of one minute is considered acceptable only if more than half of the samples in the set are left after elimination. The same rejection ratio (of 50\%) is used to ascertain the overall quality of the data for a night. Out of a total of 109 nights observed, 38 nights were of good quality. Pivot point altitudes of 3 m, 12 m, 48 m, 192 m and 768 m were chosen for their uniform logarithmic spacing. The distribution of integrated atmospheric seeing from ground level till the pivot point altitudes (taken over the two-year campaign) are given in Figure~\ref{fig:compiled_seeing}.

\begin{figure}
\centering
\includegraphics[width=1\textwidth]{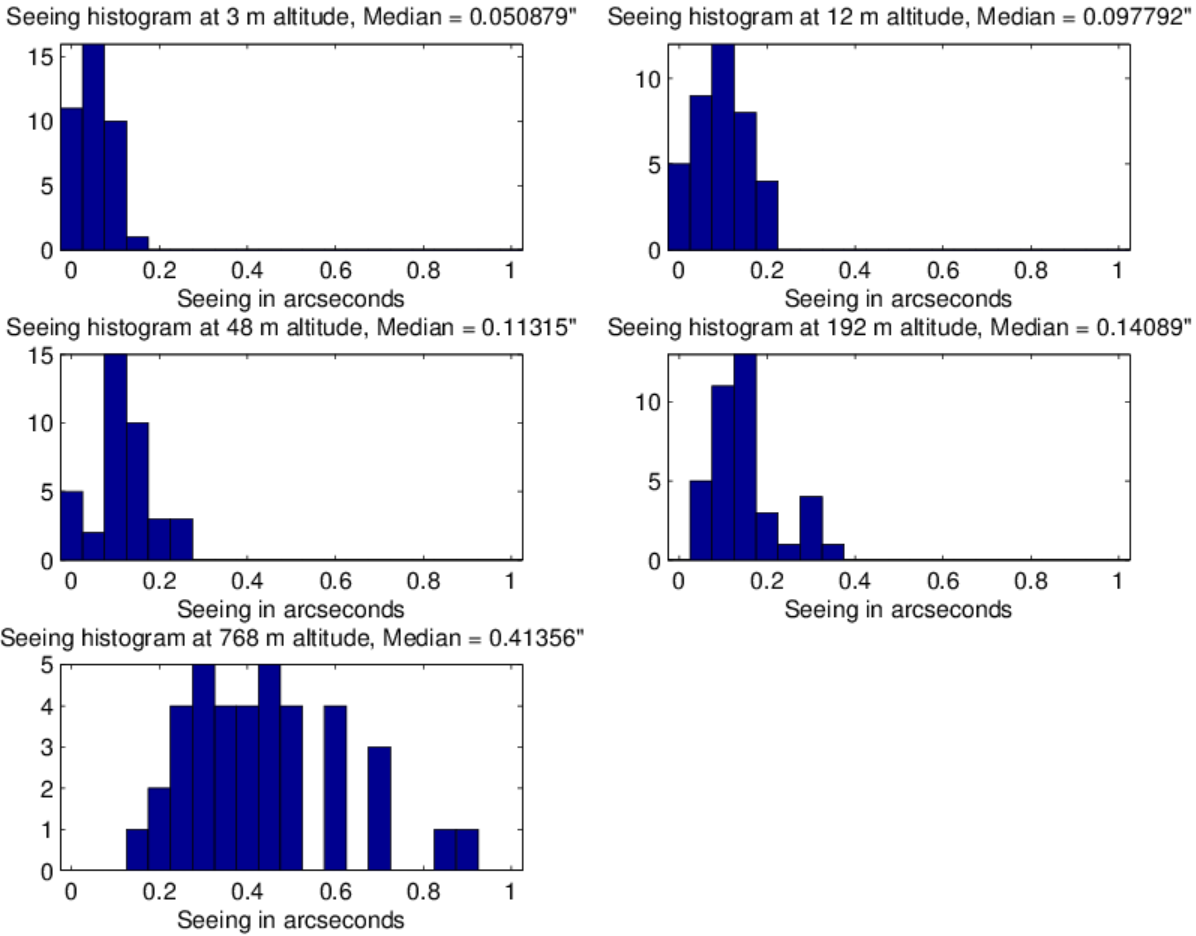}
\caption{Histogram of the integrated atmospheric seeing from the ground till the pivot points for the complete campaign. \label{fig:seeing_hist}}
\end{figure}

\begin{table}[]
\centering
\caption{Median atmospheric seeing (at a wavelength of 500 nm) and mean ${C_n^2}$ for the pivot point altitudes.}
\label{tab:seeing}
\begin{tabular}{llllll}
\hline
   & Altitude & \pbox{25cm}{Integrated \\Seeing} & ${C_n^2}$ (in m$^{\nicefrac{-2}{3}}$) at each layer &  \\
\hline
1. & 3 m      & 0.05"            & \num{4.48e-15} ${\pm}$ \num{7.47e-15}            &  \\
2. & 12 m     & 0.10"           & \num{3.34e-15}  ${\pm}$ \num{1.03e-14}            &  \\
3. & 48 m     & 0.11"           & \num{2.28e-16} ${\pm}$ \num{1.65e-15}            &  \\
4. & 192 m    & 0.14"           & \num{1.93e-16} ${\pm}$ \num{7.36e-16}            &  \\
5. & 768 m    & 0.41"           & \num{4.22e-16} ${\pm}$ \num{6.48e-16}            & \\
\hline
\end{tabular}
\end{table}

We faced some issues regarding the performance of certain channels and contamination by ambient sources in the earlier half of the two-year campaign, but all issues were resolved by mid-2014. Due to the presence of precipitation and clouds, the number of good quality data points were low between the months of March and August every year. Hence, more data points are available in the months between September to February than in the rest of the months. 

\begin{figure}
\centering
\includegraphics[width=1\textwidth]{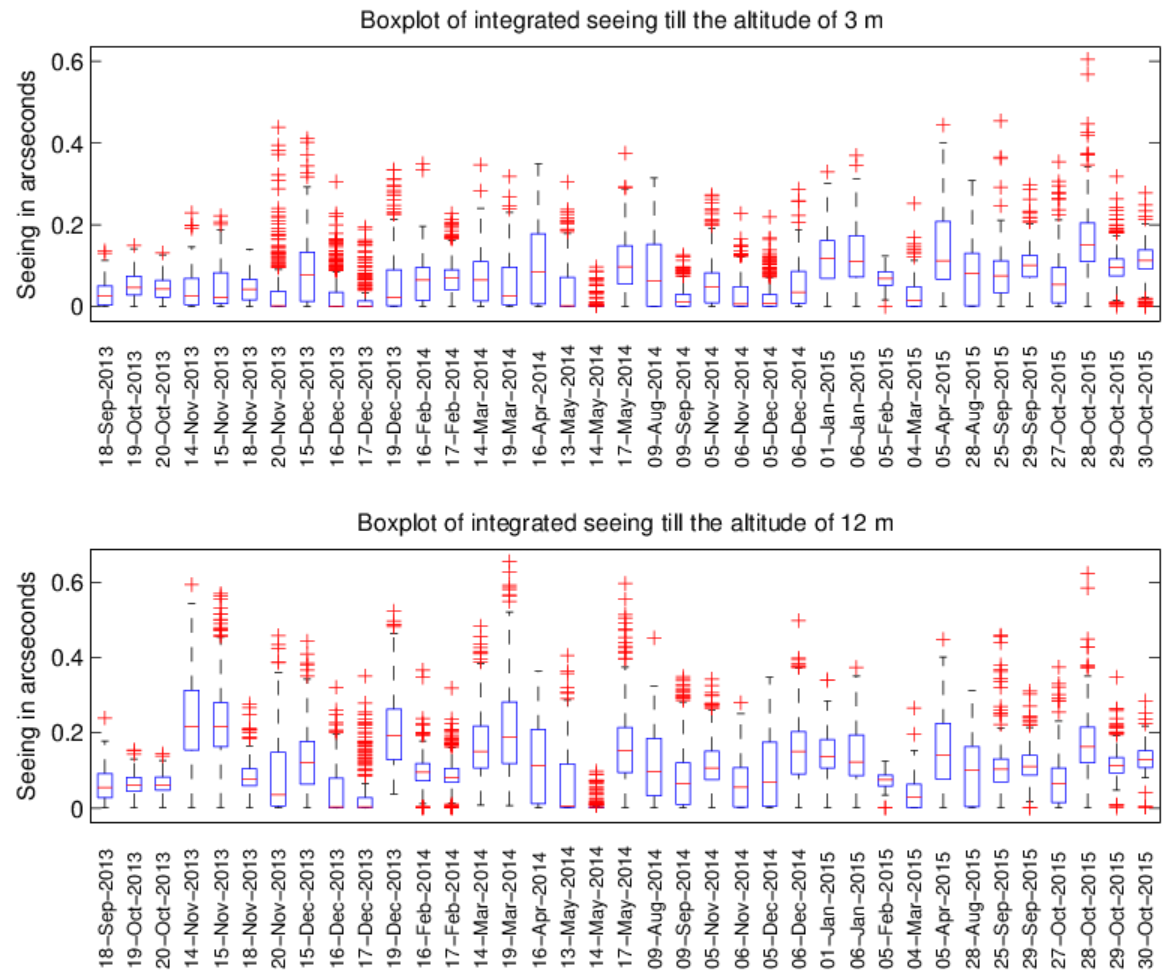}
\caption{Intranight variation of integrated seeing from ground till the altitudes of 3 m (top panel) and 12 m (bottom panel). The boxes represent the spread of the interquartile range (middle 50\% of the distribution of seeing), the bands inside the boxes represents the median seeing for each night and the red crosses represent the data points outside the interquartile range (outliers). \label{fig:boxplot1}}
\end{figure}

\begin{figure}
\centering
\includegraphics[width=1\textwidth]{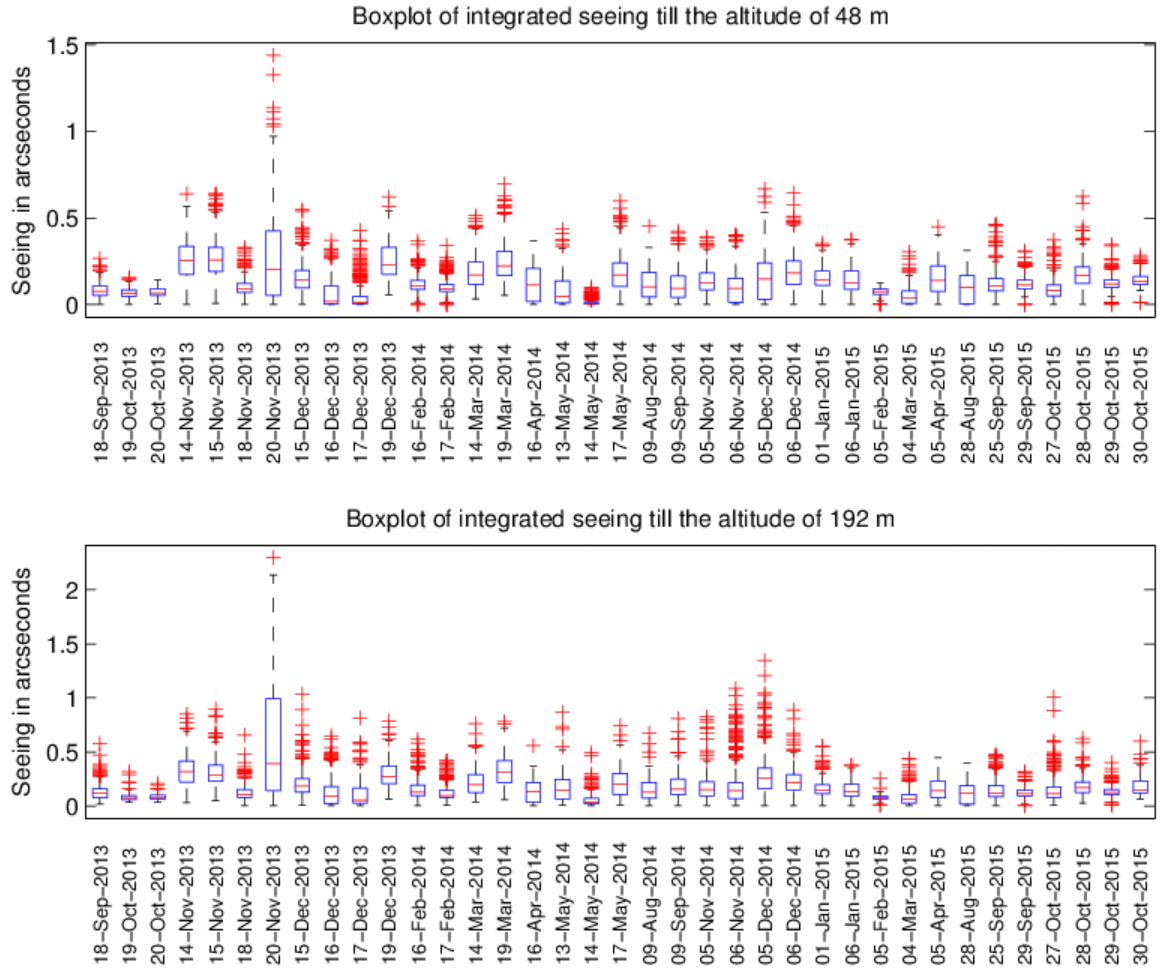}
\caption{Intranight variation of integrated seeing from ground till the altitudes of 48 m (top panel) and 192 m (bottom panel). The boxes represent the spread of the interquartile range (middle 50\% of the distribution of seeing), the bands inside the boxes represents the median seeing for each night and the red crosses represent the data points outside the interquartile range (outliers). \label{fig:boxplot2}}
\end{figure}

\begin{figure}
\centering
\includegraphics[width=1\textwidth]{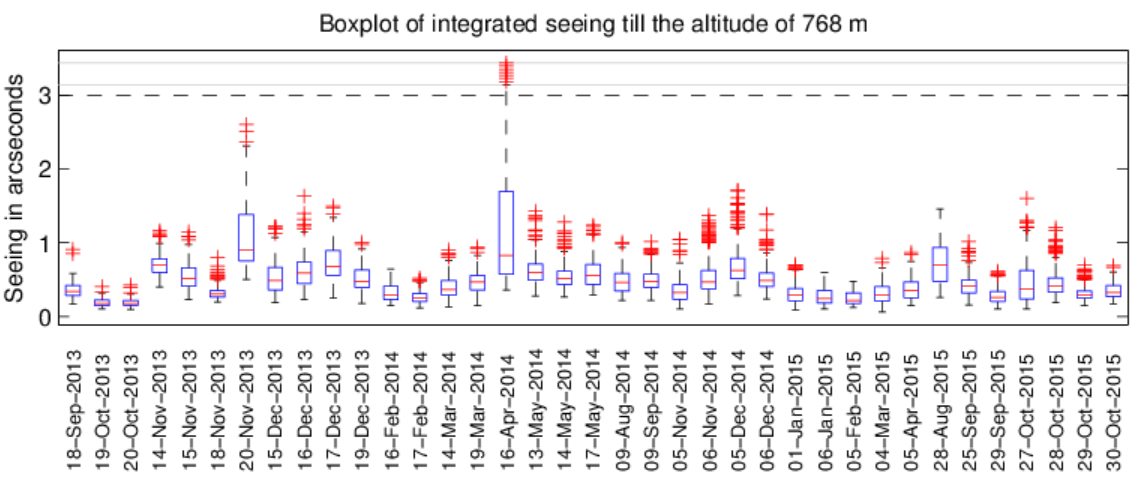}
\caption{Intranight variation of integrated seeing from ground till the altitudes of 768 m. The boxes represent the spread of the interquartile range (middle 50\% of the distribution of seeing), the bands inside the boxes represents the median seeing for each night and the red crosses represent the data points outside the interquartile range (outliers). \label{fig:boxplot3}}
\end{figure}

The overall site turbulence characterized by the ${C_n^2}$ and integrated atmospheric seeing values at the five pivot points are given in Table~\ref{tab:seeing}. The integrated seeing is the atmospheric seeing from ground level till the pivot point altitude. The histogram of integrated seeing till the different pivot point altitudes are given in Figure~\ref{fig:seeing_hist}. The average value of ${C_n^2}$ at the altitudes of 3 m and 12 m are found to be lower compared to the measurements obtained from other prominent astronomical sites using micro-thermal fluctuations \citep{vernin1992, vernin1994, pant1999} and through Sound Detection and Ranging \citep{fumihiro2008}. The average ${C_n^2}$ is highest at the lowest pivot point of 3 m, and the value reduces till the altitude of 192 m. The limitation on the value of the longest baseline (40 cm) between the photodiodes leads to a reduction in the accuracy of measurement of the ${C_n^2}$ value at an altitude of 768 m. However, a similar phenomenon was noticed by \citet{goodwin2013} at the Siding Spring observatory, where the surface layer turbulence initially decreases till a certain altitude and then increases. The upper boundary layer might be subjected to a non-laminar flow of wind created by the nearby taller mountain peaks which surround the site at which IAO is located. This could explain the higher turbulence values at the altitude of 768 m (assuming the reliability of measurements at the altitude of 768 m to a certain degree). The lowest layer of 3 m is measured to contain the maximum value of ${C_n^2}$ on most of the nights on average, due to local convection. However, there have been some nights where the ${C_n^2}$ value at 12 m was found to be greater than the ${C_n^2}$ at the altitude of 3 m. Plots for every night similar to what is shown in the figures under Section~\ref{subsec:SAM} can be made available, but are outside the scope or length of this paper.

The boxplots of integrated atmospheric seeing for the different pivot points are shown in Figure~\ref{fig:boxplot1} -~\ref{fig:boxplot3}. The data is represented in the form of a boxplot to show the dispersion and skewness of atmospheric seeing conditions for the different nights. The intranight variation in atmospheric seeing is greater between the months of March and August. The ground-layer seeing is found to be on the higher side during winter season (November-January). Astronomers who observe at HCT have consistently complained of poor atmospheric seeing in winter, and the atmospheric seeing values from DIMM \citep{parihar2015} point to the same.

The median seeing at the IAO site at Hanle from DIMM measurements is found to be 1.27" \citep{parihar2015}, whereas the median atmospheric seeing measured by the lunar scintillometer for its highest pivot point altitude is 0.41" (Table~\ref{tab:seeing}). The turbulence integral till the pivot point altitude closest to the boundary layer contributes to just 11\% of the total value. The same ratio of the boundary layer atmospheric seeing to the total seeing is 35-60\% at  Paranal \citep{lombardi2008}, 60\% at Mauna Kea \citep{roddier1990, skidmore2009} and 63\% at La Palma \citep{vernin1992}. The contribution of ground-layer seeing to total seeing is less at the IAO site when compared to other prominent astronomical sites around the world. 

Although the lunar scintillometer measures atmospheric turbulence through a direct optical method and is self-calibrated, further work is to be done to simultaneously observe the turbulence with conventional site survey instruments (like MASS and DIMM) and compare their results with the same. The work has demonstrated the robustness of the instrument and multiple versions will be fabricated to analyze the earmarked sites for the NLOT. The non-intrusiveness of the instrument is advantageous compared to microthermal towers and masts. We also plan to enable the instrument for online data processing and robotic control in the future.

\begin{acknowledgements}
The authors would like to thank Prof. Andrei Tokovinin for his valuable correspondence and advice. We would like to thank the staff at IAO-Hanle, notably Mr. Dorje Angchuk and Mr. Sonam Jorphali. We would also like to thank Mr. Urgyan Dorjey, Mr. Dadul and Mr. Phunchok Angchok for the endless nights spent in monitoring the working of the instrument and the prompt despatch of data from the site. We are thankful to the staff at the IIA mechanical workshop including Mr. P.M.M. Kemkar and Mr. Periyanayagam for the mechanical design and fabrication of the instrument and the LX200 piggyback mount.  We are also thankful to the Indian Institute of Astrophysics and the Department of Science and Technology for funding this project.
\end{acknowledgements}

\appendix

\section{Computation of the signal-to-noise ratio due to photon noise} \label{app:fmoon}
The extremely low signal strength requires us to ascertain the contribution of photon noise and the associated signal-to-noise ratio. Important parameters like the photocurrent generated by the full moon and the signal output due to the weakest of scintillations are also computed in this section. 

\begin{figure}
\centering
\includegraphics[width=0.8\textwidth]{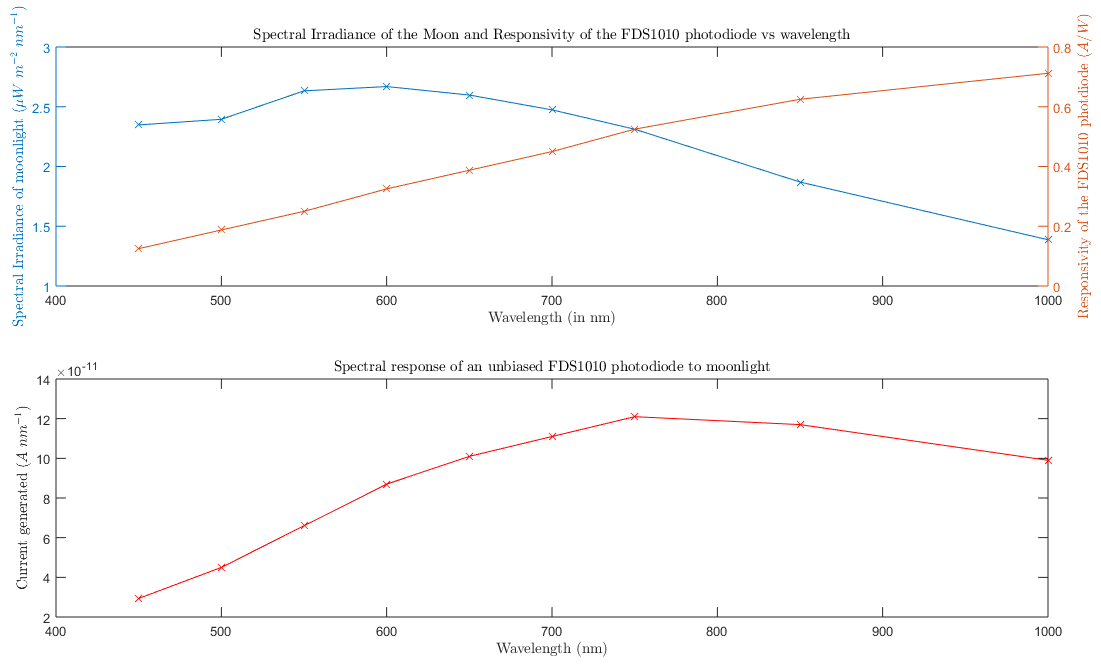}
\caption{Upper plot shows the spectral irradiance of moonlight (in $\mu W\ m^{-2}\ nm^{-1}$) and the responsivity of the FDS1010 photdiode (in $A/W$). The lower plot shows the spectral response of the photodiode to moonlight (in $A\ nm^{-1}$) \label{fig:spec_irr}}
\end{figure}

\cite{cramer2013} has created a spectral irradiance profile of the moon with a combined standard uncertainty of less than 1\%, between the wavelengths of 420 nm to 1000 nm, unaffected by strong molecular absorption. Although the differences in the locations of observation and the corresponding sky conditions will result in a difference in the observed spectral irradiance, we can use the spectral response generated in \cite{cramer2013} to estimate a rough value of the signal-to-noise ratio at the lower limits of the covariance of normalized intensity fluctuations (close to $10^{-4}$). The upper plot in Figure~\ref{fig:spec_irr} shows the spectral irradiance of moonlight (the data for which is derived from \cite{cramer2013}) in units of $\mu W\ m^{-2}\ nm^{-1}$, and the responsivity of the FDS1010 photdiode (the data for which is derived from the FDS1010 datasheet \citep{fds1010}) in units of $A/W$. The lower plot in Figure~\ref{fig:spec_irr} shows the spectral response of the photodiode to moonlight, in units of $A\ nm^{-1}$, which is the product of the spectral irradiance of moonlight with the responsivity of the FDS1010 photodiode (both of which are given in the upper plot of Figure~\ref{fig:spec_irr}) and the area of the photodiode ($10^{-4}\ m^2$). The area under the curve shown in the lower plot (of Figure~\ref{fig:spec_irr}) would give an approximation of the current generated by the unbiased FDS1010 photodiode when exposed to the full moon at zenith. The area under the curve gives a current of $I_{dc}=52.36\ nA$. The spectral irradiance data given in \cite{cramer2013} only covers the wavelength range of 420 - 1000 nm. Hence, the actual current generated will be higher than the calculated DC photocurrent ($I_{dc}$) due to the full moon, if we include the photoelectrons generated for the wavelengths at which the data is not available. For the sake of computing the signal-to-noise ratio, we have taken the $I_{dc}$, which is computed for the area under the curve shown in the lower plot of Figure~\ref{fig:spec_irr}.

For an integration time of $t = 2\ ms$, the total photoelectrons generated by the direct current,
\begin{eqnarray} 
N_s&=& \frac{I_{dc} \times t}{e}=\frac{52.36 \times 10^{-9} \times 0.002}{1.6 \times 10^{-19}}\nonumber \\
&=&6.55 \times 10^{8} \label{eq:ns}
\end{eqnarray}

The number of photoelectrons contributing to photon noise is given by,
\begin{eqnarray}
\sigma&=&\sqrt{N_s}=2.56 \times 10^{4}, \\ \label{eq:sigma}
I_\sigma&=&\frac{\sigma \times e}{t}=2.05 \times 10^{-12}A
\end{eqnarray}

The output voltage contribution of the photon noise (after amplification) is given by,
\begin{eqnarray}
V_{o\sigma}&=&I_\sigma \times K_{trans} \times K_{ac}\nonumber \\
&=&2.05 \times 10^{-12} \times 9.1 \times 10^{6} \Omega \times 90 \nonumber \\
&=&1.68 mV,
\end{eqnarray}
where $K_{trans}$ is the gain of the transimpedance amplifier and $K_{ac}$ is the AC gain of the inverting amplifier.

The amplitude of scintillations can be as low as $10^{-4}$ times the average flux amplitude of the moon. From equation \ref{eq:ns}, the number of photoelectrons generated by the smallest of scintillations can be given by,
\begin{eqnarray}
N_{sc}&=&N_s \times 10^{-4} \nonumber \\
&=&6.55 \times 10^{4} \label{eq:nsc}
\end{eqnarray}

From equations \ref{eq:sigma} and \ref{eq:nsc}, for the lowest range of scintillation amplitudes, the signal-to-noise ratio (SNR) because of photon noise is given by
\begin{equation}
S_p=\frac{N_{sc}}{\sigma}=3.2
\end{equation}

\bibliography{LuSci}{}

\end{document}